# Cooperative spontaneous emission induced by smaller metal nanoparticles

George J. Vathakkattil, [1] M. Praveena, [2] J. K. Basu, [2] and Murugesan Venkatapathi [1]*

[1] *Computational & Statistical Physics Laboratory, Department of Computational & Data Science, Indian Institute of Science, Bangalore, 560012*

[2] *Department of Physics, Indian Institute of Science, Bangalore, 560012*

*\*murugesh@serc.iisc.ernet.in*

We present a study showing cooperative behavior of light emitting quantum dots at *room temperature*, with large increases in radiative decay rates and efficiencies, in the presence of small gold nanoparticles (1.5 - 4 nm radii) in low fractions. This is a size-regime of metal particles where the expected effect on emission from independent emitters is vain non-radiative loss. But the addition of such metal particles in low fractions induces a strong evolution of the super-radiant modes of emission among quantum dots and aids their survival of thermal fluctuations; exhibiting a phase transition. While an increase of size of metal particles results in an increase in local thermal fluctuations to revert to the behavior of apparently independent emitters. Our theoretical evaluations of their possible collective modes of emission in the presence of metal nanoparticles predict such experimental observations.

Two different types of self-assembled nanoscale structures containing quantum dots were experimentally studied. This included the effect of the fractions and size of metal particles on the collective modes of emission in each type of structure; each type of structure had samples of different nominal sizes (and emission energies) of dots to establish generality. First, quantum dots collected in cylindrical cavities surrounded by randomly located gold particles were experimentally studied in large ensembles using polymer templates. The other type of nanostructure was a colloidal monolayer of quantum dots closely packed along with small gold nanoparticles. A cross-over between collective and independent regimes is observed based on the size of metal particles, and also at larger number fractions in the closely packed structure. Time-resolved photoluminescence measurements were also used to confirm this increase in the quantum efficiency and radiative decay rates of the dots.

The *spontaneous* emission from $N$ strongly interacting emitters can be very different from their independent behavior, and a coherence that increases the probability of spontaneous emission (and radiated power) by $N^2$ is called the Dicke effect.[1] Although Dicke's effect of spontaneous emission has been observed in cooled gases and mirror coupled atoms,[2-5] emitters in a bulk material cooperatively interacting with each other to result in coherent emission has been elusive. On the other hand, a related mechanism of light emission i.e. the coherent *stimulated* emission is ubiquitous in its application as a laser. The potential applications have motivated many theoretical studies on super-radiant/coherent spontaneous emission in ensembles of emitters.[6-10] The splitting of a mode of emission into many eigenstates of an ensemble of emitters can give rise to a rich collective behavior, but it is seldom observed in bulk materials. Experimental indicators of cooperative emission with long range interactions between quantum dots at very low temperatures were reported.[11] Thermal effects on collective emission are also not well elucidated, though it is expected to be a relatively weak dependence when fluctuations in vacuum are the only concern.[5] In homogeneous ensembles of emitters, a balanced mixture of super-radiant and other sub-radiant (inhibited) modes of emission preclude any large scale effect of the collective behavior that is apparent. This mixture of modes along with the effect of thermal fluctuations allows one to use independent-excitation models widely and effectively to most of the emission phenomena observed. Meanwhile, cooperative decay of emitters into a metal surface was suggested many decades ago and again recently, using theoretical arguments.[12, 13] In general, interactions of the emitters with other matter renders the problem of evaluating their collective modes difficult. Other recent theoretical works showed that the emitters around a single metal nanoparticle in the center also include super-radiant modes in the long-wavelength limit; while cooperative quenching by the metal particle for the closely placed emitters was also predicted.[14, 15] Collective modes of emission evolve due to exchange of virtual photons, but in presence of these metal particles, involve virtual interactions of plasmons as well. But the limitation of a single metal particle, and a very large number of emitters around it, was too restrictive in comparison to experiments where many metal particles may be involved. The other important unknown variables in this case are the size of metal particles, mechanism of evolution of modes, and associated thermal sensitivity.

More significant were the experimental observations; a variation of radiative decay rates of a set of emitters moved away in increments from a small gold nanoparticle (~ 6nm radii), without any apparent effect on its non-radiative losses,[16] contradicting the behavior of independent emitters interacting with a metal particle. Other experimental observations by us displayed similar characteristics as well,[17,18] and were suggested to be the effect of collective modes of emission.[19,20] A computational method to evaluate the collective self-energies and eigenstates of a set of emitters interacting with other matter entities in general heterogeneous mixtures, without the restriction of long-wavelength approximations, was also reported.[21] A limitation of using long-wavelength approximations to evaluate collective effects of emitters around a metal particle was that the sub-radiant modes could not be evaluated. We note that some sub-radiant modes can play a vital role in dissipating thermal fluctuations through the metal particles. In this work, we used such evaluations to predict the evolution of collective eigenstates, and more importantly, show the survival of these collective modes against thermal fluctuations in bulk materials. These results predicted observables of our experimental measurements with multiple types of Cadmium Selenide (CdSe) quantum dots and nano structured films with the small gold nanoparticles; which contradict the behavior of independent emission from quantum dots.

The radiative decay rates and quantum yield of emitters such as atoms, fluorescing molecules and quantum dots can also be altered by *significant* changes in proximal matter.[22-25] This is due to a change in local density of optical states (LDOS) around the emission energy - otherwise called as Purcell effect on the *independent* emitter. Experimentally, metal nanoparticle systems have so far been studied to enhance the emission process by Purcell effect, and concentrate the effect of the excitation radiation using strong near-fields, both due to the localized surface *plasmon* resonances in the metal particles. This implied they consist of metal nanoparticles of relatively large sizes, as the net gain in emission is proportional to the scattering efficiency of these nanoparticles, to a first approximation. A larger metal particle also allows larger concentration of excitation radiation in its near-field, which is proportional to the sum of its scattering and absorption efficiencies; here efficiency is given by the corresponding energy normalized by cross-sectional area of the particle. In this size regime, the absorption in the metal particle increases as $\sim a^3$ while scattering increases as $\sim a^6$ from a negligible value, which has been the primary factor in using larger nanoparticles of dimensions comparable to the wavelengths of emission and excitation (where radius $a$ of particles used is typically $50 - 100$ nm).[26, 27] Such larger metal particles involve non-radiative losses as well, though the increase in radiative rates is relatively larger and hence there is a gain in quantum efficiency when the emission is close to the plasmon resonance. The work described here on the other hand, involve metal particles that are an order of magnitude smaller in sizes.

From an application point of view, super-radiant emission as demonstrated by us, can increase efficiency of emission and also reduce non-radiative thermal dissipation. Significantly increased radiative rates reduce the material volume required to emit a desired power. The exploitation of such systems as coherent sources of light through interaction between larger numbers of emitters seems possible in principle. This principle of induced coherence in spontaneous emission may be even more useful in sources of higher energy radiation, where coherent stimulated emission has practical difficulties. In the next section, we first describe a general mechanism where coherence emerges in a quantum system due to its interactions with a dissipative resonance such as a plasmon in metal particles.

**Emergence of coherence in a quantum system due to virtual interactions with a dissipative resonance:**

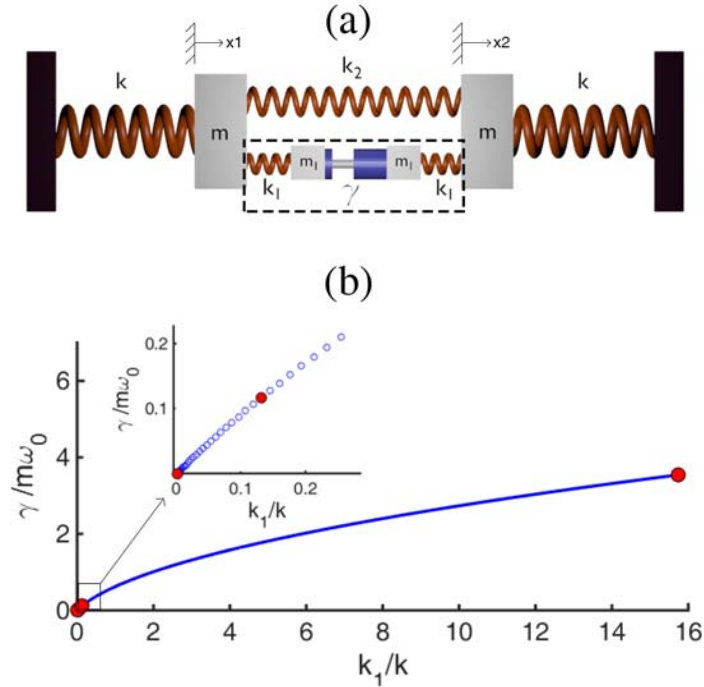

Figure 1: (a) Mechanical model of two excitations coupled by a damped resonance that is marked by a box with doted lines; $k_2 << k$ and $x_1, x_2$ are the displacements of masses $m$ (b) The parametric space of $k_1$, $\gamma$ studied as in Figure-2b and red markers point to 3 parametric points represented in Figure-2a.

We begin with a simple mechanical model and its quantum interpretation to elucidate the strong emergence of coherence among a set of excitations when they are coupled through a dissipative resonance. Such a mechanical model of two excitations allows us to briefly ignore the more algebraic task of evaluating actual interactions in a quantum system, and instead, highlight features of a mechanism relevant to many systems in generality. Figure-1a shows two mechanical oscillators (denoted by spring constant $K$ and mass $M$) coupled through a dissipative system in the center. Let the dissipative system have its resonance ($\omega_d = \sqrt{k_1/m_1 - \gamma^2/4}$) relatively close to the resonance of the oscillators ($\omega_o = \sqrt{k/m}$). The two relevant parameters to be studied here are (1) emergence of coherence between oscillators for random initial conditions and (2) the dependence of this coherence behavior with the 'size' ($k_1/k$, $m_1/m$) of the dissipative system with the relatively stationary resonance ($\omega_d \sim \omega_o$). Figure-1b shows the parametric variation of the dissipative system which approximately represents the varying absorption and extinction of plasmonic metal particles of increasing size; though we note that any other physically realistic choice of this function would not distort generality of the conclusions to other quantum systems. Figure 2a highlights the coherence between the two oscillators for random initial velocities for three different sizes of this dissipative system; these points are marked red in fig-1b. Note that the relationship between the two oscillators is incoherent when the dissipative resonance is absent. But then coherence emerges when a dissipative system of appropriate size is introduced, and a further increase in the size of the dissipative system results in the loss of coherence.

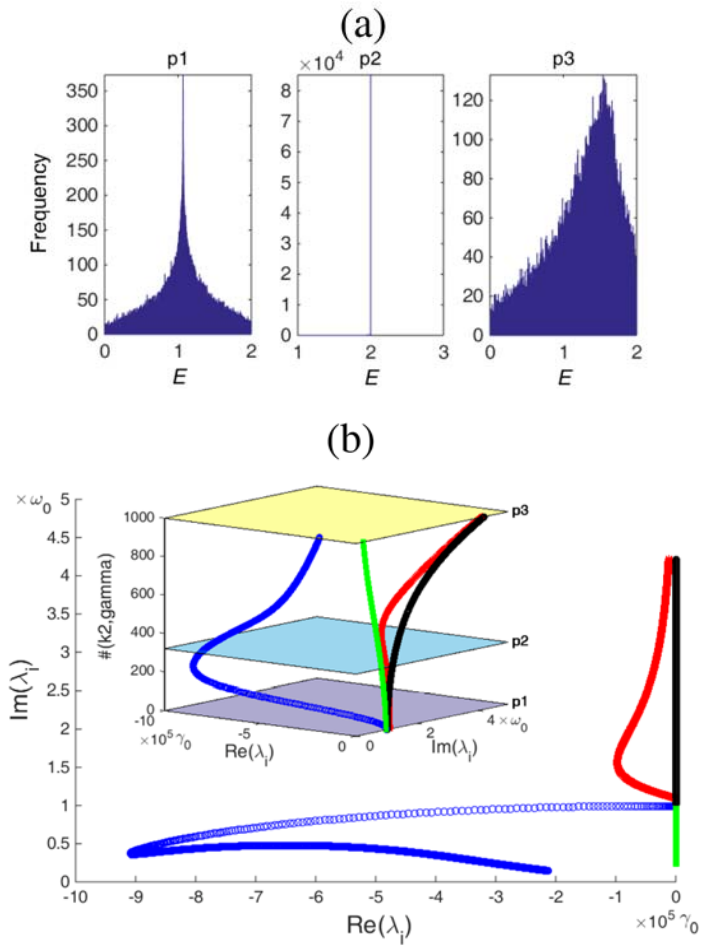

Figure 2: (a) Coherence described by $E = (x_1 + x_2)^2 / [x_1^2 + x_2^2]$ at the 3 points of the parametric space marked red in Figure-1b: $10^5$ time samples using $10^3$ random initial conditions from a uniform distribution (b) Traces of eigenvalues for all parametric points in Figure-1b

To elucidate the mechanism involved, we first plot the traces of the eigenvalues of the four modes representing this system with varying size of the dissipative system (in Fig-2b). Firstly, we discuss two of these modes that involve relatively smaller changes in the resonance frequency of the oscillators, which are given by the imaginary part of eigenvalues. One mode (in green) involves the coherent motion of all the four masses with a negligible dissipation and this represents the coherent super-radiant modes in this analogy. The other mode (in blue) includes both, the pair of masses $m$ of the oscillators and the pair of masses $m_1$ of the dissipative system 180° out of phase between them, and a 90° phase difference with masses $m$ and $m_1$. This sub-radiant mode quickly removes any incoherent motion between the oscillators, and this energy is dissipated in the system representing the metal particles. As the size of dissipative system increases further, this mode becomes ineffective and approaches that of independent oscillators with a purely imaginary eigenvalue. We will begin quantum interpretations of this system after a discussion of the other two modes. The other two modes (in red and black) represent cases where masses can be considered as sets of hybrid masses $m + m_1$ with higher energy internal oscillations. The two hybrid masses can be coherent (black) or out of phase (red) and these modes involve large energy shifts represented by the strongly varying imaginary parts of their eigenvalues. Such hybrid excitations of combined excitons and plasmons have been recently named as plexcitons.[28,29]

In the above mechanical model, the emerging coherence was at the cost of energy lost to the dissipative system, which need not be the case in quantum systems. The phase relationship and frequency of oscillations of the above mechanical model can also represent coherence and energy of two excitations interacting with each other. Energy of photon represents the energy of radiative transition between two energy levels, and the time evolution of the Hamiltonian is given by the self-energies.[30] In the optical regime, radiation damping is small ($\gamma_o << \omega_o$), and in the limit of weak coupling with vacuum, Lorentz dipoles as quantum oscillators can be used to represent such excitations for evaluating the interactions with each other and with metal particles. In the quantum model, the decay of the oscillators and the energies of interactions represent probabilities; interactions through metal particles can be explained as exchanges or annihilation of virtual plasmons in the coherent and the incoherent modes respectively. Also interactions are much shorter in lifetime than the emission process and are on the order of the lifetime of plasmons; highlighted here by real part of the eigenvalues of damped modes (in Figure 2b) that are much larger than radiation damping $\gamma_o$.

The strength of dissipative modes in removing incoherence is proportional to the absorption efficiencies of the particles. A local cyclic energy balance is maintained in the virtual interactions through the incoherent energy removed and the random phase thermal fluctuations of a surrounding equilibrium; and thus coherence emerges. As the size of metal particles increases, scattering and absorption efficiencies increase resulting in strong independent interactions of each emitter with the metal particles, and also an increase in thermal fluctuations in the vicinity of the metal particle. Only the former effect was captured in the mechanical model but the latter further reduces the size of metal particles that can play a role in inducing coherence. A larger scattering efficiency results in increase in thermal fluctuations near metal particles across the energy spectrum,[31] reducing induced coherence of emitters in the bulk material. Also, increased fluctuations in the vacuum around the metal particle result in higher rates of independent spontaneous emission from emitters; the Purcell effect. This makes the regimes of Purcell effect and induced cooperation mutually exclusive, and the latter is limited to metal particles < 10 nm radii where scattering cross-section is negligible (see Figure S6 for the sensitivity of this effect to temperature and size of particles). Unlike the long-wavelength limit where 3 super-radiant modes and $N$-3 sub-radiant modes exist for $N$ emitters, such a sharp classification of modes is not valid in general, especially in inhomogeneous mixtures. Also, the phase relationship of emitters in the super-radiant modes can be more complex due to location and polarization of oscillators in three dimensions, and a number of metal particles. Evaluations of self-energy matrices and radiative properties of many emitters interacting with each other and metal particles are compared with our experiments in the next section.

**Results and Discussion:**

The first study we present is on a colloidal single (mono) layer of quantum dots dispersed along with gold (Au) nanoparticles ~ 1.75 nm radii (Figure-3a). Photoluminescence (PL) of two different types of CdSe quantum dots was studied: (1) 1.5 nm radii with 2.35 eV peak emission energy and (2) 2.6 nm radii with 2.05 eV peak emission energy. Each of these were studied in two different packing densities and this gave us four different types of samples for any metal fraction (see Table-1 in SI for full details). In these monolayer films the metal surface-dot separations are 1.5 – 4.5 nm depending on the sample. The raw data of PL intensity spectra shows an increase up to ~ 5 times due to the addition of small gold nanoparticles in the lower fractions and it reduces as the fractions of metal particles increase (Figure-S5 in SI). The radiative rates increase up to 3 times in the time-resolved PL measurements, and the quantum efficiency calculated is presented in Figure-3b here. Firstly, such large increases in radiative rates or improved quantum efficiencies are unexpected, as these smaller metal nanoparticles in the monolayer result in absorption, and also energy transfer loss from a neighboring quantum dot (when distance from the surface of a metal particle < 3 nm). Thus the radiative rates are expected to remain unperturbed for smaller fractions, but non-radiative losses should increase with addition of metal particles and this should result in a sharp decrease in efficiency. Figure-3b highlights this contradiction where the expected quantum efficiency of independent dots is included.

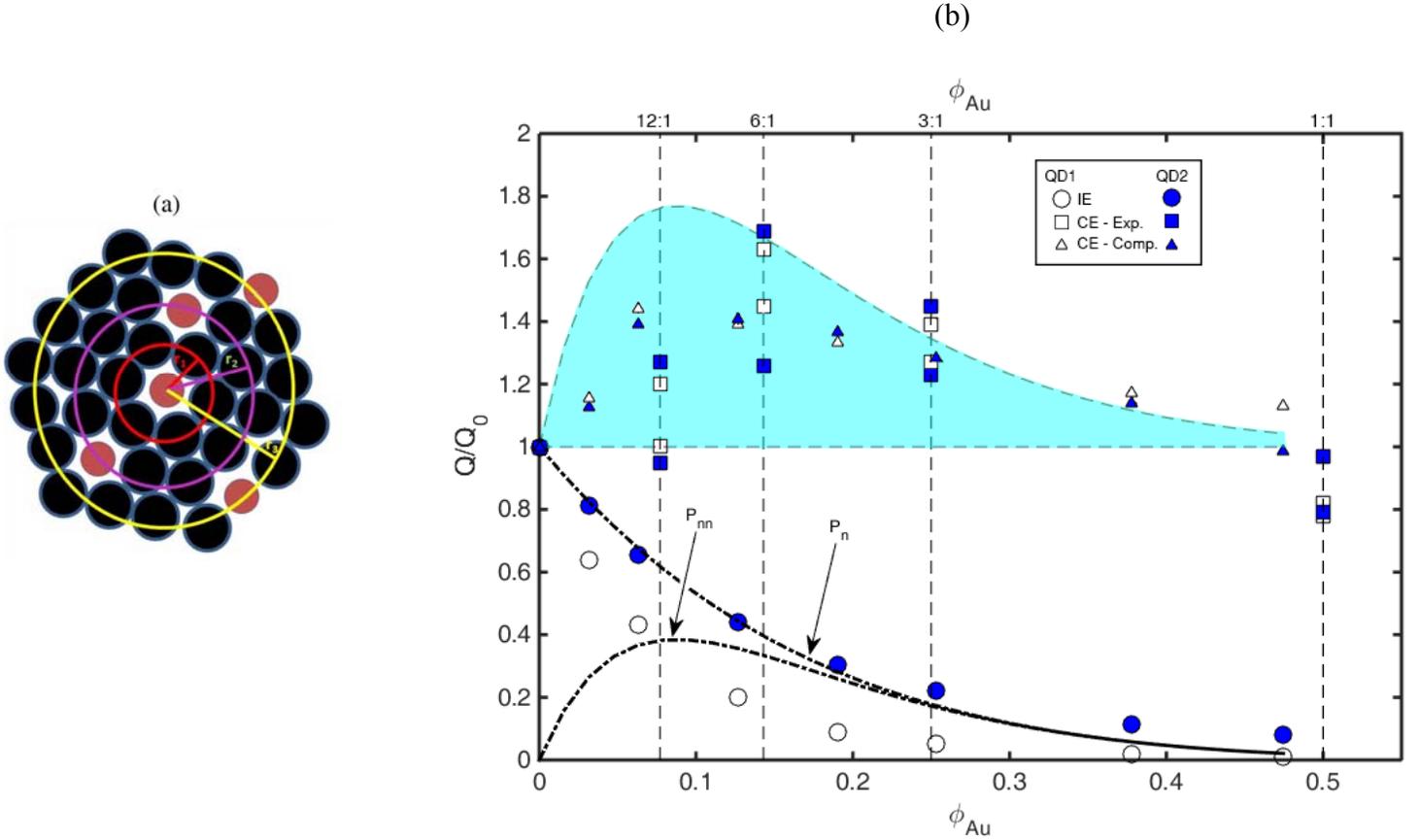

Figure 3: (a) A sketch of monolayer, Au particles in red circles and QDs in black (b) Quantum efficiency measured and calculated for various cases. QD1: 2.35 eV peak emission energy and QD2: 2.05 eV peak emission energy. Each fraction for a QD has two packing densities; detailed structural parameters and corresponding Q values provided in Table S1. **CE**: Collective Emission; **IE**: Independent Emission (expected); **Exp.** and **Comp.** refer to experiments and computations respectively. The envelope predicts the increase of efficiencies using the ideal long-wavelength superradiance efficiency of 3 for a fraction of dots represented by $P_{nn}$ in the monolayer, using relation $1 + 2P_{nn}$ as the upper-bound.

But large radiative decay rates and improved quantum efficiencies are indeed predicted by the model of collective emission where gold nanoparticles induce a strong evolution of the highly radiant modes. The model seems to marginally under-predict both collective emission and quenching effects in the film, but the trends are unmistakable. The effect of the gold particles even on the emission energies far away from its plasmon resonance at 2.38 eV is notable. The small metal particles split collective modes into two groups; one with large radiative rates and the other include virtual dissipative modes. Note that large gains in radiative rates manifest when the number fraction of metal particles is ~ 0.15, and further increase in their fractions reduces the emission to less than the baseline values. This is because metal particles neighboring a quantum dot while playing a crucial role in splitting the collective modes, also reduce the quantum efficiency of the highly radiant modes. For these dots, off-diagonal terms of non-radiative decay matrix are large and comparable to its diagonal terms.

To reduce this into a first approximation, in Figure-3b we plot the probability of a dot to 'not' have a metal particle neighbor in the monolayer. Assuming 6 first-neighbor vacancies and 12 second-neighbor vacancies as in hexagonal arrangement of this monolayer, allows us to calculate probabilities. These probability numbers are very weakly dependent on the number of neighbor sites we assume, for example, numbers 5 and 11 are equally useful. The probability of not having a first-neighbor metal particle, $P_n$, indeed strongly correlates to the residual quantum efficiency of dots exhibiting independent behavior. In this collective emission close to plasmon resonance, efficiency seems to have a strong correlation to $P_{nn}$, the probability that a quantum dot 'does not have a metal particle as first-neighbor but does have one as a second-neighbor' which peaks at a fraction 0.12. Though it is unphysical to conclude second-neighbors play a much larger role than first-neighbor metal particles, it has a deeper physical explanation of the balanced role played by metal neighbors as explained above. More importantly, increasing the size of gold nanoparticles to 4 nm radii makes this effect vanish; these samples show no gain in radiative rates even for the low fractions around 0.15, and rather show significant quenching for higher fractions (Figure-4), as expected in independent emission from the dots.[23,32] For dots with emission (at 1.88 eV) far away from plasmon resonance, large enhancements of PL seems to be possible with the larger gold particles of 4nm radii in this film,[34] but at higher fractions ~ 0.5. This agrees with expectation that induced cooperation is possible when both sufficient absorption required for virtual dissipative modes and a very low scattering by metal particles is manifest.

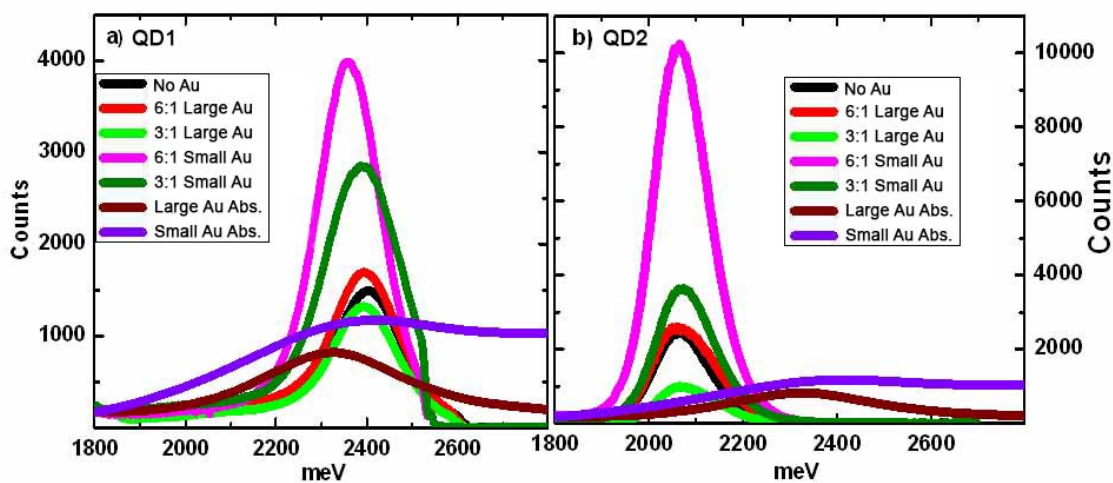

Figure 4: Photoluminescence intensity spectra for Au fractions 6:1 and 3:1. **Large Au** corresponds to samples with gold particles of 4 nm nominal radii and **Small Au** to 1.75 nm. The absorption curves of Au particles have arbitrary units not included in this graph, and are a guide to the location of the plasmon resonance peaks.

We proceed to another three-dimensional self-assembled nanostructure which exhibits this induced cooperative emission, where that conclusion is even more apparent. Here the dots are well separated (ranging 10 to 50 nm) from the gold nanoparticles of ~ 2.5 nm mean radii and there are no conflicting processes like energy transfer (Figure-5a).

(a)

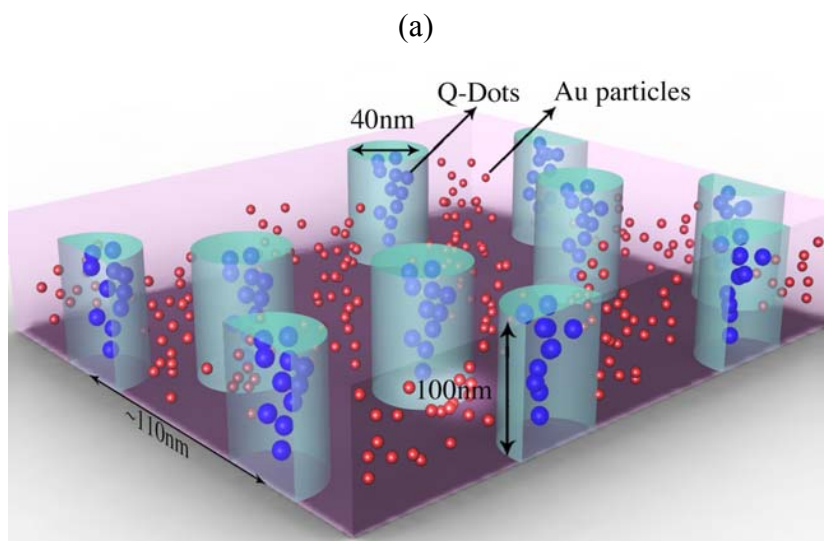

(b)

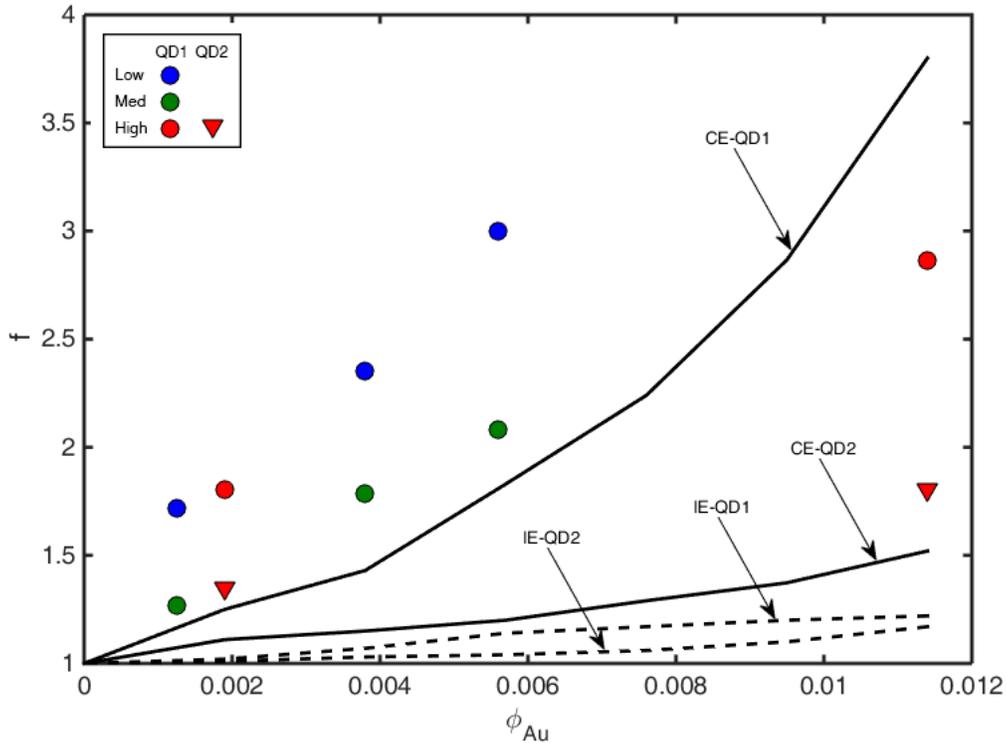

Figure 5: (a) Structure of self-assembled films and average dimensions (b) gains $f$ in photoluminescence intensity for fractions of Au; QD1: 2 nm radii with ~ 2.20 eV peak emission energy and QD2: 3 nm radii with 1.95 eV peak emission energy; Low, Medium and High correspond to 0.0135, 0.027 and 0.054 volume fractions of QDs in the film. **CE**: Collective Emission; **IE**: Independent Emission (expected).

The physical structure of these block copolymer systems and the PL observed have been described before,[17-19] and we elucidate the origin of these significant effects here. The observed enhancement in PL power and the corresponding small volume fractions of gold particles added to the film are shown in Figure-5b. Note that any configuration of such low volume fractions of gold nanoparticles, irrespective of their size does not explain the large differences in PL (given by $f$ in Figure 5b), in terms of independent behavior of the dots. Even these optimistic estimates of independent emission that ignore absorption by metal particles severely under-predict the gains. The films with a lower density of dots in the cavities exhibit large gains for even the very low fractions of gold nanoparticles. Note that a lower spatial density favors super-radiant modes as argued from a pure geometrical point of view.[33] The large enhancements of emission (at 1.95 eV), away from the weak plasmon resonances of the small gold nanoparticles even with high packing densities of QDs, is a strong evidence for a mechanism that goes beyond the negligible increase in local density of optical states.  The collective emission model predicts large enhancements observed due to the strongly radiant modes.  This model inherently assumed high spatial density of emitters inside the cylindrical cavities because of random spatial sampling, and can be considered an approximate lower-bound for the emission expected from dots.  As observed in the monolayer films, larger concentrations of gold nanoparticles may not be conducive for cooperative emission.  We have also examined similar nanostructures with an unstable size regime for metal particles, which exhibit quenching as expected from independent emitters.  For example, a sample in which the added 1 nm radii gold particles aggregate to large particles of various sizes averaging ~ 10 nm radii. [17, 19]

In summary, we presented a theoretical formalism and experiments that showed the strong evolution of super-radiant modes of spontaneous emission in the presence of smaller metal nanoparticles.  This produced observations that were 5-10 times larger than possible otherwise with the inclusion of smaller metal nanoparticles.  Sub-radiant modes seem to play an important role in inducing coherence among the emitters by virtual energy dissipation in the metal particles.  Thermal fluctuations of equilibrium maintain an energy balance in this process but when scattering efficiencies become non-negligible, this induced coherence vanishes.  This formalism explained our experimental studies with quantum dots which were outside the realm of our understanding in terms of their independent emission.  From a utility point of view, super-radiance can increase quantum efficiency by factors even greater than 2 for low efficiency emitters, and the larger radiative decay rates are significant for reducing the volume of emitting materials required.  Further studies on coherence properties of the induced super-radiant modes are pertinent, and other improvements in theoretical calculations are also possible.  This principle of induced coherence may have larger implications for higher energy sources where stimulated emission has practical difficulties, though its use in coherent sources of optical energies can not be discounted in the future.

**Methods:**

*Preparation of materials:*

CdSe QDs, capped with pyridine, were synthesized using the method developed by Peng & Peng with modification,[35] while thiol terminated polystyrene (PST) capped gold nanoparticles were prepared by chemical reduction of gold chloride in presence of PST.[36]  To prepare the hybrid arrays of QDs and gold nanoparticles, we used block copolymer templates of polystyrene-polyvinylpyridine (PS-P4VP, Polymer Source) of molecular weight 330–125 kg/mol (PS: P4VP) (C3) which is known to form hexagonal arrays of cylinders of P4VP in the PS matrix in bulk.[37, 38]

In thin films, the cylinders can be made to align perpendicular to the substrate after incorporation of CdSe QDs. The mixed solutions of pyridine capped CdSe QD, BCP, and PST capped gold nanoparticles in toluene were spin coated on glass substrates and cleaned by standard method of RCA Laboratories. The interaction between pyridine ligands at the surface of the CdSe QD and P4VP ensures that the CdSe QDs self-assembles exclusively into the P4VP cylinders while the PST capped gold nanoparticles reside inside PS region. The spin coated films were annealed at 180 °C for 48 h under a vacuum of 10-4Torr. The thickness of the films used in this work was 100-160 nm as measured using spectroscopic ellipsometer (Sentech).

Specific details of the methods used in the monolayer films and the related materials are provided in the Supplementary Information. The Octadecanethiol selenide (CdSe) QDs of diameter 3nm, 5.2nm, and Dodecanethiol (DDT) capped AuNPs of diameter 3.5nm, 8nm are labeled as QD1, QD2 and small Au, Large Au respectively.

*Measurements:*

The Confocal PL (Alpha NSOM Witec, Germany) measurement was performed using a Argon blue laser with 488nm line at room temperature (25 0C) in transmission mode. The laser light passing through the 100X objective aperture excites the sample. The emitted light is collected using high efficiency objective (Nikon NA- 1.25, magnification 100 X), and finally guided to photo multiplier tube (PMT) for light intensity image or spectrograph for spectra collection. Several spectra were collected from the films from different locations and recorded with CCD with integration time of 10 seconds each to estimate the change in the emission intensity over various locations. The time resolved TRPL lifetime measurements were done using Horiba Scientific, Model: Fluoro cube-01-NL, life time system. Pulsed picosecond laser diode light source, Excitation source (469nm) used was of pulsed duration 70ps. Repetition rate was 1MHz, TAC range: 50ns.

*Theoretical Evaluations:*

To theoretically predict the change in emission of the interacting QDs, the perturbation of the metal particles and other QDs to the self-energy of any QD has to be evaluated. This is accomplished using the interactions of Lorentz dipole quantum oscillators and other polarizable metal particles. This procedure and the decomposition of the radiative and non-radiative components have been described elsewhere.[21] Random permutations of locations and polarizations of QDs are used in the respective nominal geometries to converge to a density of collective states in the nanostructure. The thermal stability of the evolution of super-radiant modes in each permutation is determined using the non-radiative decay of its virtual dissipative modes. Each data point of the theoretical predictions use many possible collective modes and they converge to a normal distribution in the observables of the radiative decay rates and effective quantum efficiency. These evaluations are described in detail in the supplementary information.

**Supplementary Information** is linked to the online version of the paper at…

**Acknowledgments:**

We acknowledge the Department of Science and Technology (Nanomission), India for the financial support and the Advanced facility for microscopy and microanalysis, Indian Institute of Science, Bangalore for the access to TEM and TRPL measurements. M. P. acknowledges UGC, DST-Nanomission, India for the financial support.

**Author Contributions:** J. K. B and M. P. designed and performed the experiments, G. V. J. developed the mechanical model, M. V. developed the theory and numerical codes, and all authors participated in writing the manuscript.

**Author Information:**

Competing financial interests: None.

Reprints and permissions information is available at <u>...</u>

Correspondence about the theoretical methods can be addressed to <u>murugesh@serc.iisc.in</u>, and requests regarding experiments can be addressed to <u>basu@physics.iisc.ernet.in</u>.

# Cooperative spontaneous emission induced by smaller metal nanoparticles

George J. Vathakkattil, [1] M. Praveena, [2] J. K. Basu, [2] and Murugesan Venkatapathi [1]*

[1] Computational & Statistical Physics Laboratory, Department of Computational & Data Science, Indian Institute of Science, Bangalore, 560012

[2] Department of Physics, Indian Institute of Science, Bangalore, 560012

*murugesh@serc.iisc.ernet.in

***Materials & Measurements:***

The pyridine capped CdSe QDs were synthesized using the earlier method [1], and thiol terminated polystyrene PST capped gold nanoparticles were prepared described earlier [2]. The BCP templates were prepared with polystyrene-polyvinylpyridine (PS-P4VP) (C3) which is known to form hexagonal arrays of cylinders of P4VP in the PS matrix in bulk [3,4]. In the case of monolayer QD films, the Cadmium Selenide QDs with 2 different sizes were synthesized following methods described earlier [5,6]. In a typical synthesis, cadmium oxide (CdO) were dissolved in trioctylphosphine oxide (TOPO) on being heated at $310^0$C and $270^0$C, Selenium solution in trioctylphosphine (TOP) was injected rapidly to this solution. Color of the solution changed to reddish, orange indicating the quantum dots formation. The solution was kept at respective temperatures for ten minutes. The nanoparticles powder was extracted from the reaction mixture by adding methanol after cooling down to room temperature. The powder settled at the bottom of the vessel were collected and the capping of TOPO is exchanged with ODT (octadecanethiol), according to methods reported previously in the literature [7], with appropriate modifications. The 1-Dodecanethiol capped AuNPs of diameter 3.5nm was synthesized using methods described earlier [8,9]. The small volume of as prepared GNPs from the solution was dispersed in chloroform solvent (Thermo Scientific Barnstead Nanopure, 18.2 M $\Omega$) and used for UV visible absorption spectroscopy. The absorption measurements were done using Perkin-Elmer UV-Vis absorption spectrometer. Photoluminescence measurement was done using very dilute solutions of quantum dots dispersed in the chloroform. Using the known concentration of quantum dots in the solution and that of standard dye rhodamine 6G as reference, we calculated [10] the quantum yield of quantum dots. To prepare samples for TEM imaging, a small drop of respective solutions were transferred to carbon coated copper grids. The grid samples are dried for 12 hours. The size and shape of the CdSe quantum dots is characterized using transmission electron microscope (TEM) images as shown in the Figure S2 on the TEM grids for the characterization. The mean diameter of the larger CdSe QDs was 5.2±0.01 nm(QD2) while that of the smaller QDs was 3±0.02 nm(QD2).

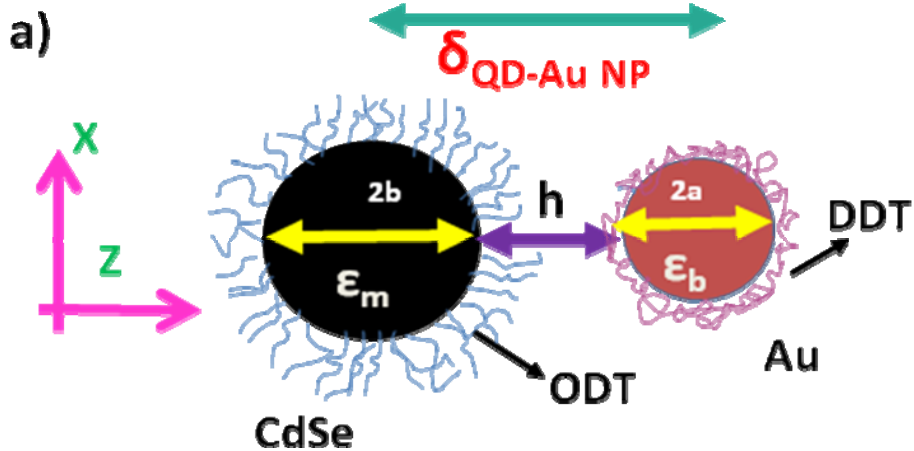

<u>Figure S1:</u> a) Schematic of the configurations of Au NPs and CdSe QDs in the CQD films. Here, **δ**=h+a and the average inter particle separation between QDs and AuNPs is **δ$_{QD-AuNP}$**. 2b and 2a are the diameter of the CdSe and AuNP core. ODT and DDT are the polymer capping ligands of the CdSe QD and Au nano particles with corresponding dielectric constants $\epsilon_m$ , $\epsilon_b$ respectively.

The bare and hybrid QD monolayer deposited on the glass substrates, in a controlled way using a Langmuir-Blodgett (KSV instruments, Finland) is described here. Cadmium selenide quantum dots and Au nanoparticles were mixed in various ratios ($N_{CdSe}/N_{Au}$) in chloroform and spread over de-ionized water-filled Langmuir trough (KSV mini trough, Finland). Sufficient waiting time was given (15 min) to evaporate the chloroform completely. The temperature of the water sub phase is maintained at constant temperature of 25 $^0$C with the help of regulated water circulation (Julabo, Germany). The surface pressure was monitored using a platinum Wilhelmy plate. The barriers were compressed and decompressed with constant speed of 10 mm/min. The typical surface pressure (∏) versus Area (*cm$^2$*) of the isotherm is recorded using the built-in-software. Langmuir monolayer was transferred over glass substrates for optical studies after cleaning using the standard RCA method. (Hydrogen peroxide, *H$_2$O$_2$*, and ammonium hydroxide (*NH$_4$OH*) were added in the boiling de-ionized water (150$^0$C) in the volumetric ratio of 1:1:5, respectively). The substrates were immersed in the boiling solution at constant temperature for 10 min and taken out, rinsed four to five times with running de-ionized water before use in Langmuir monolayer transfer. This method makes the substrates hydrophilic. The substrates were inserted before spreading the quantum dots on the water sub-phase. After the two isotherm cycles, the substrates were pulled out of the Langmuir monolayers vertically at the rate of 0.5 mm/min at a constant surface pressure (∏) of 35 to 20 mN/m for H and L density cases. The transferred film was then vacuum-dried for 12 hrs. Prepared hybrid films are represented in the schematic diagram as shown in Figure S1. The confocal measurements are performed on these monolayer films under ambient room temperature.  The further details of the calculated quantum yield and the sample details are provided in the table 1 below.

The Confocal measurements were performed on these hybrid films which were prepared with combination of no Au, small Au and large Au nano particles for 6:1 and 3:1 of QD1,QD2 cases. The photoluminescence (PL) counts are plotted along with the UV-Visible absorption of basic AuNPs to show the overlap regime of QD1 case. Whereas, the UV-Vis absorption plots are not overlapping with the QD2 PL peaks as shown in Figure S3.

**S2**

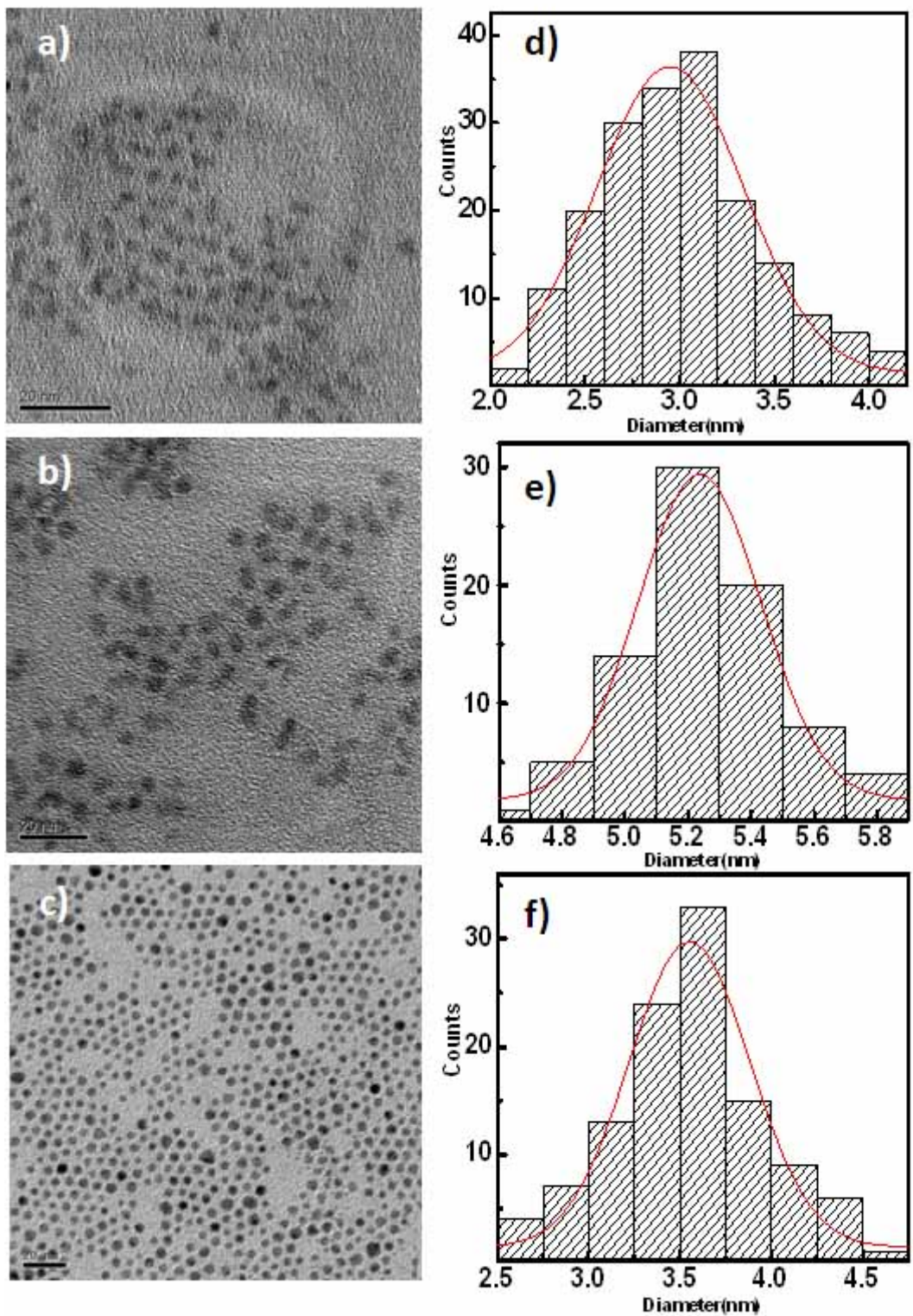

**Figure S2:** TEM images of 2 types of CdSe QDs and the AuNPs. The scale bar is 20 nm.

TABLE S1: *Details of Samples with measured quantum yield ratio from TRPL measurements*

| Sample index | $N_{CdSe} : N_{Au}$ | h (in nm) | $\rho$ (in nm$^{-2}$) | Q/Q$_0$ |
|---|---|---|---|---|
| | 1:0 | 2:555 | 3.1*10$^{-2}$ | |
| | 12:1 | 2:485 | 3.2 *10$^{-2}$ | 1.2 |
| QD1-H | 6:1 | 2:135 | 3.1 *10$^{-2}$ | 1.63 |
| | 3:1 | 1:715 | 3.2* 10$^{-2}$ | 1.39 |
| | 1:1 | (:::) | .. 10$^{-2}$ | 0.82 |
| | 1:0 | 4:725 | 1.9* 10$^{-2}$ | |
| | 12:1 | ::: | .. 10$^{-2}$ | 1.001 |
| QD1-L | 6:1 | 4:095 | 2.0* 10$^{-2}$ | 1.45 |
| | 3:1 | (:::) | .. 10$^{-2}$ | 1.27 |
| | 1:1 | (:::) | .. 10$^{-2}$ | 0.78 |
| | 1:0 | 1:715 | 3.1* 10$^{-2}$ | |
| | 12:1 | 1:645 | 3.2* 10$^{-2}$ | 1.27 |
| QD2-H | 6:1 | 1:470 | 3.2 10$^{-2}$ | 1.26 |
| | 3:1 | 1:400 | 3.3 *10$^{-2}$ | 1.23 |
| | 1:1 | (:::) | .. 10$^{-2}$ | 0.79 |
| | 1:0 | 3:220 | 2.0 *10$^{-2}$ | |
| | 12:1 | (:::) | .. 10$^{-2}$ | 0.95 |
| QD2-L | 6:1 | 2:625 | 2.0* 10$^{-2}$ | 1.69 |
| | 3:1 | (:::) | .. 10$^{-2}$ | 1.45 |
| | 1:1 | (:::) | .. 10$^{-2}$ | 0.97 |

Later, time resolved PL (TRPL) measurements were performed on these samples. TRPL measurements were done on these samples using, Horiba Scientific, Model: Fluoro cube-01-NL, life time system. The excitation source wavelength of 469 nm which is a pulse pico second laser diode of pulse duration 70 Pico seconds with a repetition rate of 1 MHz and TAC range of 50 nano seconds. A peak preset of 20,000 counts was used during data collection. The collected decay rate profiles are shown in the Figure S4.

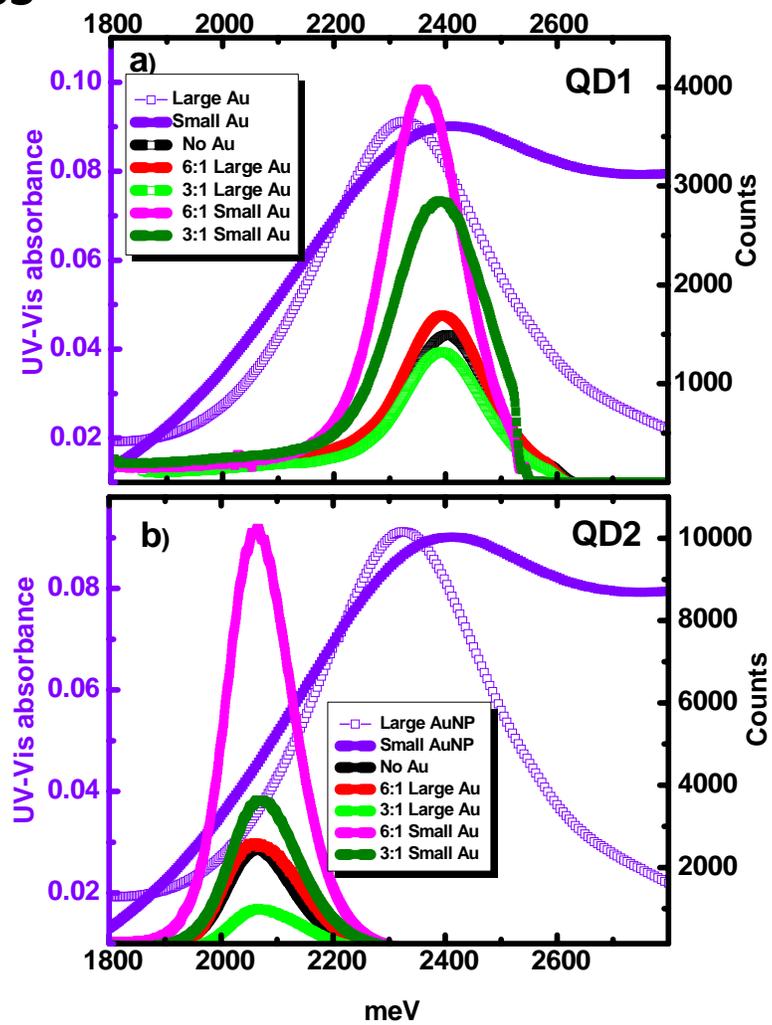

**Figure S3:** Photoluminescence intensity spectra for Au fractions 6:1 and 3:1. **Large Au** corresponds to samples with gold particles of 4 nm nominal radii and **Small Au** to 1.75 nm.

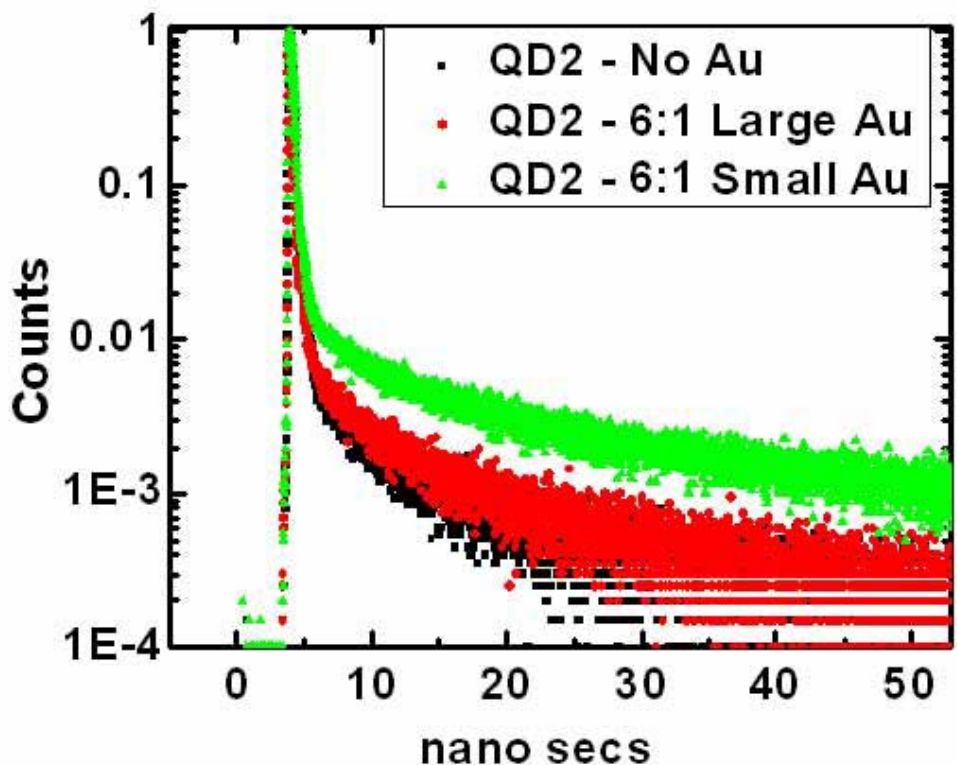

**Figure S4:** The TRPL decay profiles of the QD2 monolayer films with large and small Au nanoparticles (1.75 nm and 4 nm nominal radii respectively).

The experimental observations of PL enhancement factor $F_{Int}= (I_{hyb}/I_{QD})$, is plotted in Figure S6a, where $I_{hyb}$ is the PL intensity of CQD films doped with AuNP and $I_{QD}$ is the PL intensity of bare CQD film. The similar observations from independent TRPL measurements estimates the PL enhancement as, $F_{LT} = \Gamma^R_{hyb} / \Gamma^R_{QD}$, as shown in Figure S6b. Here, $\Gamma^R_{hyb}$ and $\Gamma^R_{QD}$ are the total radiative decay rates of the hybrid and bare QD films respectively. The ratio of the quantum yield (Q) of the hybrid and the bare samples is calculated for all the QD1-series and QD2-series and given in the table S1.

**S6**

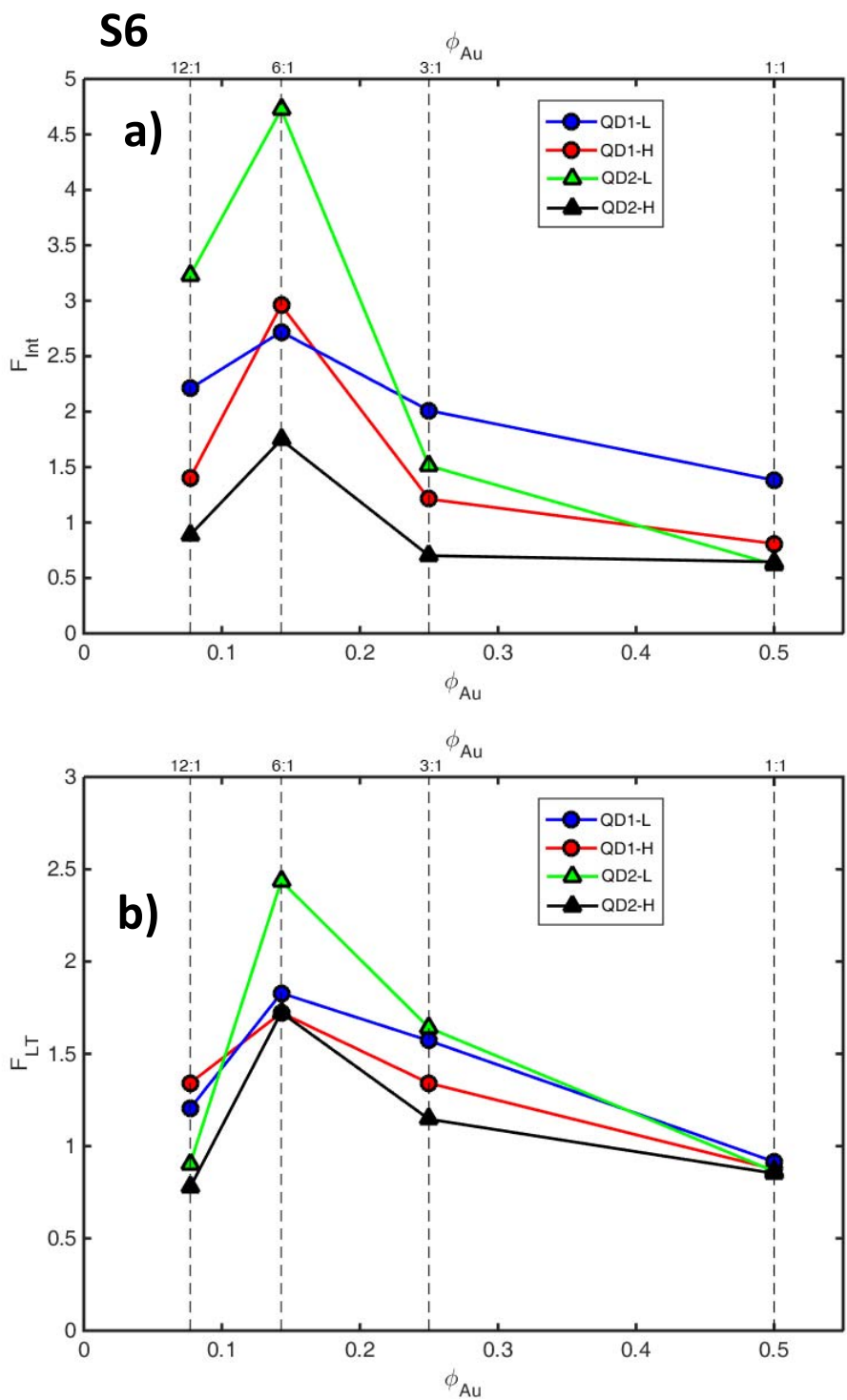

**Figure S5:** Experimental values of enhancement with respect to $F_{Au}$ in respective samples as indicated in the panels. (a) from PL intensity measurements and (b) from TRPL measurements data.

***Theoretical Evaluations:***

The measured PL spectrum of a size-dispersed sample of quantum dots can be elucidated as a convolution of the decay of independent excitations with marginally different energies [10]. The theoretical model evaluates the observables of collective emission from an ensemble of $n$ Lorentzian quantum dots with identical energies of independent emission $\sim \hbar\omega_o$, the peak/mean PL energy as observed in experiments. This should not alter the physical conclusions derived from a more cumbersome model that includes excitations in a size-dispersed sample of emitters. The time dependent decay of the interaction Hamiltonian of the excited state and the amplitude of wave functions in an isolated dot, are represented here by the radiative and non-radiative decay rates $\Gamma_o^r$ and $\Gamma_o^{nr}$; the quantum efficiency $Q_o$ given by the ratio of the radiative rate and the sum of the two rates. The coupling of each emitter to the common radiation field can be evaluated using Lorentz dipoles with a quantum of energy $\hbar\omega_o$ each (where $\hbar = h/2\pi$ and $h$ is the Planck's constant). The common radiation field need not be explicitly quantized in the limit of weak coupling with vacuum, and this allows us to evaluate the self-energies of an interacting ensemble as shown below [11]. The perturbation of the metal particles (and other dots) to the self-energy of any quantum dot has both radiative and non-radiative components to be evaluated. Given the location of the $n$ dots and the $m$ metal particles, the self-energies defined below and their radiative/non-radiative components can be calculated without using a long-wavelength approximation, as explained elsewhere [12].

$$\Sigma^r_{jk} = \Delta^r_{jk} - \frac{i\Gamma^r_{jk}}{2} = \frac{-2\pi q^2 \omega}{mc^2} \mathbf{e}_j \cdot \mathbf{G}^r(\mathbf{r}_j, \mathbf{r}_k; \omega) \cdot \mathbf{e}_k - \delta_{jk}\frac{i\Gamma^r_o}{2} \tag{1}$$

$$\Sigma^{nr}_{jk} = \Delta^{nr}_{jk} - \frac{i\Gamma^{nr}_{jk}}{2} = \frac{-2\pi q^2 \omega}{mc^2} \mathbf{e}_j \cdot \mathbf{G}^{nr}(\mathbf{r}_j, \mathbf{r}_k; \omega) \cdot \mathbf{e}_k - \delta_{jk}\frac{i\Gamma^{nr}_o}{2} \tag{2}$$

Here, any two dipole moments $P_j\mathbf{e}_j$ at $\mathbf{r}_j$ and $P_k\mathbf{e}_k$ at $\mathbf{r}_k$ are coupled by corresponding Green tensors. Each collective eigenstates $J$ of a sampled geometry is represented by energy shifts $\Delta_J$ and decay rates $\Gamma_J^r$, $\Gamma_J^{nr}$. The radiative decay rate $\Gamma_J^r$ is given by the vector-matrix-vector product $J^\dagger \Gamma^r J$ described by the state representation

$$\sum |J\rangle = \Delta_J - \frac{i\Gamma_J}{2}|J\rangle \tag{3}$$

$$\Gamma_J^r = \langle J \mid \Gamma^r \mid J \rangle \qquad\qquad\qquad (4)$$

These eigenstates are given by the eigenvalues and eigenvectors of the self-energy matrix $\sum_{jk} = \Delta_{jk} - i\Gamma_{jk} / 2$. The diagonal terms of the self-energy matrix $\Delta_{ii}$, $\Gamma_{ii}^r$ and $\Gamma_{ii}^{nr}$ give us the characteristics of the independent emission in the heterogeneous mixture, and can be used to deflate the collective effects. These diagonal terms $\Delta_{ii}$ represent the energy shifts of each dot, and the relative increase in local density of optical states is given by the increase in decay rates $(\Gamma_{ii}^r - \Gamma_o^r) / \Gamma_o^r$. Modes numbering $n$ for one random configuration of the ensemble, are represented by $\Delta_J$, $\Gamma_J$; and many such random configurations have to be evaluated to populate the phase space of the problem sufficiently. Number $n$ can be chosen such that it is numerically and physically meaningful; a small $n$ is numerically favored for matrix operations and given sufficient sampling of random geometries in a system, the collective observable should produce a normal distribution. Note that the number of metal particles is fixed by the fraction of metal particles, and this number can not be reduced by sampling. We have used the number of emitters *n = 20* for the described structures in this work as any lesser number makes the evaluations dependent on number *n*. The theoretically predicted results included 20,000 modes for every case of experiment. For the monolayer, a lattice of 600 probable sites spanning a continuous disc of dimensions comparable to λ was used approximately representing two samples classified by peak emission energy and fraction of metal particles as in Table S1. The location of the metal particles and dots were chosen randomly; $\mathbf{e}_j$ are *unit* vectors of a random uniform distribution in $\Re^3$ representing the emitter polarizations. The position vectors $\mathbf{r}_j$ of the dots represent a different random uniform distribution in $\Re^2$ of this disc region.

In the case of the block copolymer films, a cylinder of 50 nm height and 40 nm diameter was populated with Lorentz dipoles of one quantum of energy each, using position vectors $\mathbf{r}_j$ from random uniform distributions in $\Re^3$. The metal particles with a mean distance of 35 nm from the axis of cylinder, as in the actual films, were generated using position vectors $\mathbf{r}_j$ from random normal distributions in $\Re^3$. $\mathbf{e}_j$ the unit vectors of a random uniform distribution in $\Re^3$ represented the emitter polarizations. While the above produces the density of collective states using the given nominal geometric parameters, the effect of thermal fluctuations and the probabilities of survival of cooperative modes in any particular random configuration '*c*' were evaluated as shown below.

$$p_c = (1 - \exp[\frac{-\Gamma_d^{nr}}{n\Gamma_o^r} - \frac{h\Gamma_{ens}}{kT}]) / \exp[\frac{kT}{h} \times \frac{\Gamma_{ind}^r}{\Gamma_d^{nr}}] \text{ where}$$

(5)

$$\Gamma_d^{nr} = \sum_n (\Gamma_i^{nr} - \Gamma_o^{nr}), \ \Gamma_{ens} = \sum_n \Gamma_i \text{ and } \Gamma_{ind}^r = \sum_n (\Gamma_{ii}^r / \Gamma_o^r - 1)$$

The dissipative modes are virtual when they are much faster than $n\Gamma_o^r$ which represents the probability of independent emission of a photon from *any* of the $n$ dots. Also, the total virtual dissipation (quantum fluctuation) in the ensemble, $h\Gamma^{nr}$, has to be larger compared to the thermal fluctuations in equilibrium, for survival of the collective modes. These two terms are represented by the numerator of the probability function in this formalism that does not explicitly include the vacuum. The increase in thermal fluctuations due to the metal particles is represented by the increase in the local density of optical states; included by the exponent as denominator of the function. The radiative rate of cooperative emission in a random configuration is given by the rates and probability of emission from any of the $n$ collective modes, as below in Eq (6).

$$\Gamma_c^r = \frac{1}{Q_{ens}} \sum_n Q_i \Gamma_i^r \text{ where } Q_{ens} = \sum_n Q_i$$

(6)

Beyond rigorous calculations in any particular geometry, it is insightful to use the first-approximation of cross-sections of a single gold nanoparticle when irradiated by an emitter very far away, and observe the size and temperature effects on the probability. This underestimates cross-sections and hence can overestimate the size regime of particles, but one could infer the sensitivities from the figure below knowing that actual experiments may require smaller metal particles and may involve more than one metal particle.

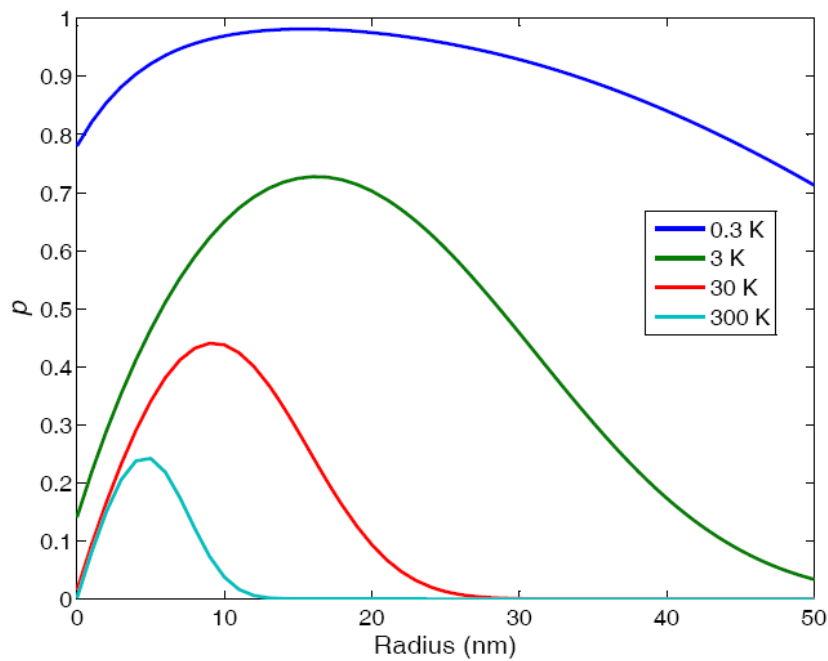

**Figure S6:** Sensitivity of the induced cooperative effect to the size of a gold nanoparticle and temperature. '*p*' refers to the estimate of probability of cooperative modes of emission using Eq (5) and the far-field limit of cross-sections of a single gold nanoparticle. The energy of emission is assumed to be 2.2eV and ε = 2.25. *p* is nearly independent of the number of emitters as can be inferred from Eq. (5).